\documentclass[12pt]{article}

\usepackage{epsf,epsfig,psfig}

\def\spose#1{\hbox to 0pt{#1\hss}}
\def\ltapprox{\mathrel{\spose{\lower 3pt\hbox{$\mathchar"218$}}
 \raise 2.0pt\hbox{$\mathchar"13C$}}}
\def\gtapprox{\mathrel{\spose{\lower 3pt\hbox{$\mathchar"218$}}
 \raise 2.0pt\hbox{$\mathchar"13E$}}}

\begin{document}
\begin{titlepage}
\begin{center}
{\Large\bf 
Confining strings in representations with common $n$-ality
}
\end{center}
\vskip 1.3cm
\centerline{
Luigi Del Debbio, $^a$
Haralambos Panagopoulos, $^b$
Ettore Vicari $\,^a$}

\vskip 0.4cm
\centerline{\sl  $^a$ Dipartimento di Fisica dell'Universit\`a di Pisa,
                       and INFN Pisa, Italy}
\centerline{\sl  $^b$ 
Department of Physics, University of Cyprus,
                      Nicosia CY-1678, Cyprus
} 

\vskip 1.cm

\begin{abstract}

We study the spectrum of confining strings in SU(3) pure gauge theory,
by means of lattice Monte Carlo simulations, using torelon operators
in different representations of the gauge group.  Our results provide
direct evidence that the string spectrum is according to predictions
based on $n$-ality. Torelon correlations in the rank-2 symmetric
channel appear to be well reproduced by a two-exponential picture, in
which the lowest state is given by the fundamental string
$\sigma_1=\sigma$, the heavier string state is such that the ratio
$\sigma_2/\sigma_1$ is approximately given by the Casimir ratio 
$C_{\rm sym}/C_{\rm f} = 5/2$, and the torelon
has a much smaller overlap with the lighter fundamental string state.

\end{abstract}

\end{titlepage}


The spectrum of confining strings in 4-$d$ SU($N$) gauge theories has been
much investigated recently.
Several numerical studies in the context of a lattice formulation of the theory
have appeared in the literature,
providing results for color sources associated with
representations higher than the 
fundamental, see e.g. Refs.~\cite{DFGO-96,Deldar-00,Bali-00,SS-00,LT-01,DPRV-02}.
General arguments show that the string tension  must depend only on the 
$n$-ality, $k={\rm mod}(l, N)$, of a  representation built out of
the (anti-)symmetrized tensor
product of $l$ copies of the fundamental representation.
The confining string with $n$-ality $k$ is usually
called  $k$-string, and $\sigma_k$ is the corresponding string tension.
Using charge conjugation, $\sigma_k=\sigma_{N-k}$.
As a consequence, $SU(3)$ has
only one independent string tension determining the large distance
behavior of the potential for $k\ne 0$.
One must consider larger values of $N$
to look for distinct $k$-strings.
Lattice results for $N=4,5,6$ \cite{DPRV-02,LT-01} 
show a nontrivial spectrum for the $k$-strings. 
In particular the data of Ref.~\cite{DPRV-02}, obtained
using color sources in the antisymmetric representations of rank $k$,  
turn out to be well reproduced  by the sine formula
\begin{equation}
{\sigma_k\over \sigma} \approx { \sin (k\pi/N) \over  \sin (\pi /N)},
\end{equation}
($\sigma \equiv \sigma_1$)
within their errors.
The sine formula 
has been suggested by several theoretical works, especially in the context of
supersymmetric theories, 
see e.g. Refs.~\cite{Strassler-98,AS-03a} and references therein.

On the other hand, numerical results for different representations
with the same $n$-ality apparently contradict
the picture that $n$-ality is what really matters.
For example, in the SU(3) case, Monte Carlo data 
for the Wilson loops for several representations
\cite{Deldar-00, Bali-00} show apparently area laws
up to rather large distances, approximately 1 fm, also
for representations with zero $n$-ality, and the extracted
string tensions turn out to be consistent with
the so-called Casimir scaling \cite{DFGO-96}.
In the lattice study of Ref.~\cite{LT-01,DPRV-02},
considering larger values of $N$,
the $k$-string tensions were extracted from the torelon masses,
i.e. from the exponential decay of correlations
of characters of Polyakov lines. 
In Ref.~\cite{DPRV-02},
while the antisymmetric representations provided
rather clean measurements of $\sigma_k$ reproducing
the sine formula, the numerical results for the symmetric representations
suggested different values of the corresponding string tensions.
For example, in the case of rank 2, $\sigma_{\rm sym}/\sigma\gtapprox 2$,
which is approximately the value suggested by 
Casimir scaling
or by the propagation of two noninteracting fundamental strings.
These results that apparently contradict $n$-ality
have been recently discussed in Ref.~\cite{AS-03b}.
They have been explained by arguing 
that standard color sources, such as Wilson loops and Polyakov lines, 
associated with representations
different from the antisymmetric ones have
very small overlap with the stable $k$-string states, being
suppressed by powers of $1/N^2$ in the large-$N$ limit,
and in some cases also exponentially.
Since $N=3$ is supposed to be already large, these arguments may explain 
why the predictions of $n$-ality
have not been directly observed in the numerical simulations,
which are limited in accuracy. Moreover, 
the situation worsens in the case of larger $N$.

Motivated by this recent work, 
we decided to return on this issue 
in the context of the  4-$d$ SU(3) gauge theory.
Performing Monte Carlo simulations of the SU(3)
gauge theory in its Wilson lattice formulation,
we measure correlators of Polyakov lines
in the representations of rank 1 (fundamental)
and 2 of the SU(3) group, in order to check  whether
their large-distance behaviors, and therefore the values of the
corresponding string tensions, are consistent  with $n$-ality.

In our numerical study we use a method based on 
torelon correlators \cite{DSST-85}.
The string tensions are extracted from the large-time behavior of ``wall-wall''
correlators of Polyakov loops in spatial directions, closed
through periodic boundary conditions (see e.g.
Refs.\cite{DSST-85,LT-01}):
\begin{equation}
G_r(t) = \sum_{x_1,x_2} \langle \chi_r [ P(0,0;0) ]\; 
\chi_r [ P(x_1,x_2;t) ] \rangle,
\label{corr}
\end{equation}
where 
$P(x_1,x_2;t) = \Pi_{x_3} U_3(x_1,x_2,x_3;t)$.
$U({\bf x}; t)$ are the usual link variables, and $\chi_r$ is the
character of the representation $r$: 
$\chi_{\rm f}[ P ] = {\rm Tr}\, P$ for the fundamental 
representation, while the two representations of rank 2
(antisymmetric and symmetric, both with $n$-ality $k=2$) 
have  
$\chi_{\rm asym}[ P ] = {1\over 2}\,(({\rm Tr}\, P)^2 - {\rm Tr} \,P^2)$,
$\chi_{\rm sym}[ P ] = {1\over 2}\,(({\rm Tr}\, P)^2 + {\rm Tr} \,P^2)$, respectively.
Note that, since for SU(3) $\chi_{\rm asym}[U]=\chi_{\rm f}[U]$,
the correlators in the $k=1$ fundamental and $k=2$ antisymmetric 
representations are identical.
We have also studied the adjoint representation ($n$-ality $k=0$), for which:
$\chi_{\rm adj}[ P ] = | {\rm Tr}\, P |^2 - 1$.
In this case,
one should consider the connected correlator, since
$\langle \chi_{\rm adj}[P] \rangle\neq 0$.

The correlators (\ref{corr})
decay exponentially as $\exp (- m_k t)$, where $m_k$ is the
mass of the lightest state in the corresponding representation.
Actually, on a finite lattice with periodic boundary conditions
we have $G_r(t)\propto {\rm cosh} (t-T/2)$, whete $T$ is the 
temporal size.
For a $k$-loop of size $L$, the $k$-string tension is obtained
using the relation \cite{DSST-85}
\begin{equation}
m_k = \sigma_k L - {\pi\over 3 L}.
\label{mks}
\end{equation}
The last term in Eq.~(\ref{mks}) is conjectured to be a universal
correction, and it is related to the universal critical behavior of
the flux excitations described by a free bosonic string 
\cite{LSW-80}.

We present results obtained at $\beta=5.9$ and for
two asymmetric lattices $12^3\times 24$ and $16^3\times 24$,
allowing us to compare the results using Polyakov lines
with different length, i.e. $L=12,16$.
Using our data for the fundamental string tension, see below,
and the standard value $\sqrt{\sigma}=440 \, {\rm MeV}$,
$L=12,16$ correspond to approximately 1.5 and 2 fm,
respectively.
In our simulations we upgraded the SU(3) variables
by alternating microcanonical over-relaxation and 
heat bath steps,  typically in a 4:1 ratio.  
More details on the algorithm can be found in Ref.~\cite{DPRV-02}.
We collected rather high statistics, $\approx 16$M
sweeps (considering a sweep as the upgrading
of all links of the lattice independently of the algorithm)
for $L=12$ and $\approx 7$M sweeps for $L=16$.
Measurements were taken every 20 sweeps.
In order to improve the efficiency of the measurements we used
smearing and blocking procedures (see e.g. Refs.~\cite{smearing}) to
construct new operators with a better overlap with the lightest string
state. 
We constructed new super-links using
three smearing, and a few blocking steps,
according to the value of $L$, i.e. two
for $L=12$ and four for $L=16$. These
super-links were used to compute improved Polyakov lines. 
Our implementation of smearing and blocking is as
follows~\cite{DPRV-02}:
Smearing replaces every spatial link on
the lattice according to:
\begin{equation}
U_k(x) \mapsto {\cal P} \left\{ U_k(x) + \alpha_s \sum_{\pm(j\neq k)} U_j(x)
U_k(x+\hat j) U^\dagger_j(x+\hat k) \right\}
\end{equation}
where ${\cal P}$ indicates the projection onto $SU(N)$ and the sum
only runs on spatial directions. Similarly, blocking replaces each
spatial link with a super-link of 
length $2a$:
\begin{equation}
U_k(x) \mapsto {\cal P} \left\{ U_k(x) U_k(x+\hat k) + 
\alpha_f \sum_{\pm(j\neq k)} U_j(x)
U_k(x+\hat j) U_k(x+\hat j+\hat k)) U^\dagger_j(x+2\hat k) \right\}
\end{equation}
The blocking procedure can then be iterated $n$ times to produce
super-links of length $2^n a$. The coefficients $\alpha_s$ and
$\alpha_f$ can be adjusted to optimise the efficiency of the
procedure. We constructed new super-links using
$\alpha_s=\alpha_f=0.5$\,.

\begin{figure}[tb]
\centerline{\psfig{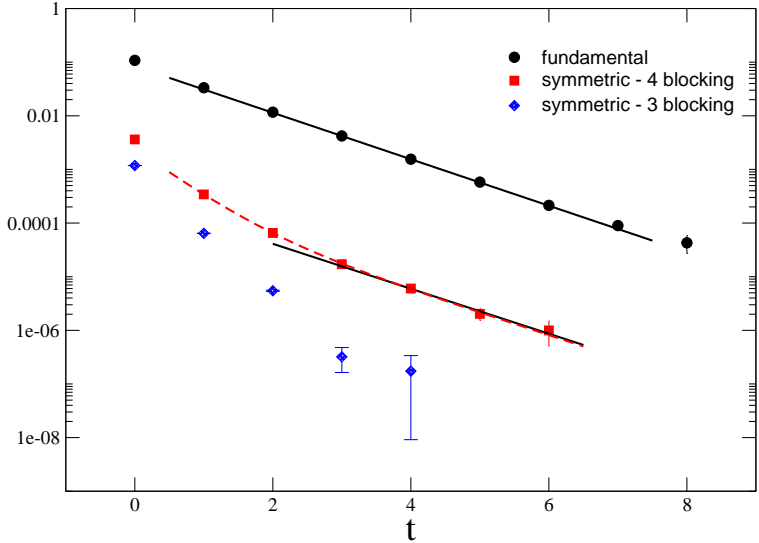}}
\vspace{2mm}
\caption{Correlator data for the fundamental (circles) and rank-2
  symmetric (squares) representations, with $L=16$, as a function of
  the temporal distance $t$.
  The top solid line is a hyperbolic cosine fit to
  data with $t\geq 3$, leading to an estimate of $m_1$\,; the bottom
  solid line is a hyperbolic cosine with the {\it same} exponent, which is well
  supported by the symmetric representation data for $t\geq 3$.
 The dashed line represents a fit using the two-exponent Ansatz (\ref{expa}). 
 Data from the rank-2 symmetric
  representation at one less blocking step is shown in diamonds.
}
\label{fig16}
\end{figure}

In Figs.~\ref{fig16} and {\ref{fig12} we show
the wall-wall correlators as a function of the distance
$t$ in the cases of fundamental and symmetric
representations, from the runs with $L=16$ and $L=12$ respectively.
The data for the correlator in the fundamental representation allow us to 
accurately determine the fundamental string tension, and the
two lattices provide consistent results using 
Eq.~(\ref{mks}), i.e.
$\sigma=0.0664(5)$ and $\sigma=0.0668(3)$ respectively
from the  $L=16$ and $L=12$ runs
(obtained by fitting results starting from distances $t=3,4$, respectively).
On the other hand, such an agreement is not observed
in the case of the symmetric representation.
However, the $L=16$ data for the symmetric correlator
shows a clear evidence that its
asymptotic behavior is controlled by the
fundamental string; indeed, fitting the data for $t\geq 3$
we obtain $\sigma_{\rm sym}=0.070(4)$,
in agreement with $n$-ality.
Although data at small $t$, $t<3$, show a clear
contamination by heavier states, 
in the symmetric representation case the overlap with the fundamental string state
of the source operator,
obtained by performing four blocking steps after smearing, appears to be sufficient
to show the actual asymptotic behavior before the signal
disappears within the error.
This is not observed in the $L=16$ data using the source operator
with three blocking steps (one less) and in the $L=12$ data
(where two blocking steps are employed).
Up to the distances that we can observe before the signals die off
into the noise, the correlators appear to be dominated by the
propagation of a much heavier state, which would suggest 
$\sigma_{\rm sym}\approx 0.16$, whose corresponding ratio
$\sigma_{\rm sym}/\sigma\approx 2.4$ is
rather close to the Casimir ratio of the two representations, i.e. 5/2.

\begin{figure}[tb]
\centerline{\psfig{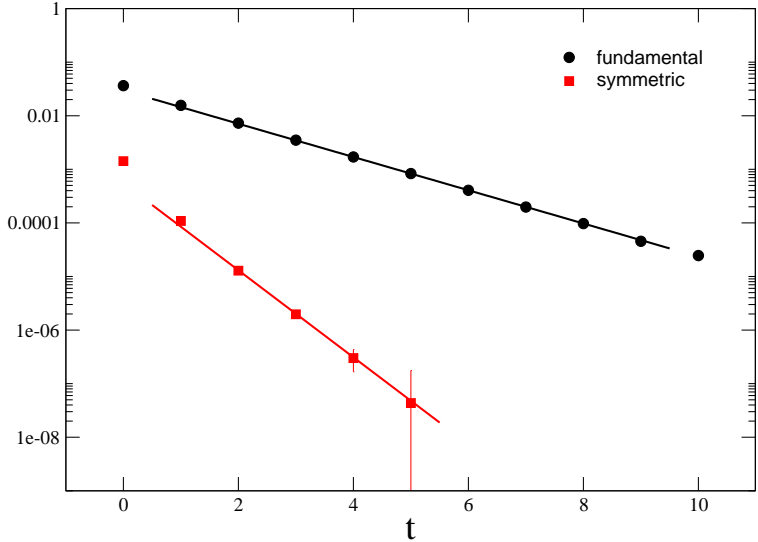}}
\vspace{2mm}
\caption{Correlator data for the fundamental (circles) and 
  symmetric (squares) representations, with $L=12$, as a function of
  the temporal distance $t$. Solid  lines are exponential fits to
  data with $t\geq 3$, leading to markedly different values of 
  $m_k$\,.
}
\label{fig12}
\end{figure}

A simple  interpretation of the behaviour of torelon correlations is
provided by a picture based on two propagating states; 
indeed, the numerical results suggest:
\begin{equation}
G_{\rm sym}(t) \simeq  c_1 e^{-m_1 t} +c_2 e^{-m_2 t} 
\label{expa}
\end{equation}
where $m_i = \sigma_i L - \pi/(3L)$. Due to $n$-ality,
$\sigma_1=\sigma$. The string tension of the first
excited string state is $\sigma_2$, and the overlap with the states
satisfy $c_1 \ll c_2$. The data on the smaller
lattice $L=12$ already suggests that the ratio $\sigma_2/\sigma$
should be approximately equal to the Casimir ratio.  As shown in
Fig.~\ref{fig16}, the Ansatz~(\ref{expa}) fits well the $L=16$ data at four
blocking steps for $t\geq 1$, with $\sigma_2/\sigma\simeq 2.2$ and
$c_1/c_2\simeq 0.12$.  Assuming a mild dependence of $c_i$ on $L$, and
given the smallness of $c_1$, the signal from the fundamental string
should only be visible for large enough $t$:
\begin{equation}
t \simeq {{\rm ln} (c_2/c_1)\over (\sigma_2-\sigma_1)\,L}
\end{equation}
Clearly, for an acceptable signal-to-noise ratio, $t$ must not grow
excessively, and the above equation requires a sufficiently large
spatial size $L$\,; this is consistent with the results from the
lattice sizes we use.

According to the large-$N$ arguments of Ref.~\cite{AS-03b},
the ratio $c_1/c_2$ should be suppressed at least by a power
of $1/N^2$. 
This would make the observation of the
asymptotic $k=2$ string state,
using sources in the symmetric representation,
much harder for larger $N$, thereby explaining the
results for $N=4,6$ of Ref.~\cite{DPRV-02}.

In conclusion, the results that we have presented provide direct
evidence that the spectrum of confining strings is according to
predictions based on $n$-ality. Torelon correlations in the rank-2 symmetric
channel appear to be well reproduced by a two-exponential picture, in
which the lowest state is given by the fundamental string
$\sigma_1=\sigma$, the heavier string state is such that the ratio
$\sigma_2/\sigma_1$ is approximately given by the Casimir ratio 
$C_{\rm sym}/C_{\rm f} = 5/2$, and the torelon
has a much smaller overlap with the lighter fundamental string state.

In the case of the adjoint representation,
since its $n$-ality is zero, the corresponding 
mass of the exponential decay in the connected correlator
should not depend on $L$, but it should be related to 
the propagation of gluelumps. 
Actually, since the gluelumps have a limited physical size,
by increasing $L$ we expect a smaller and smaller overlap
of the adjoint Polyakov lines with the lowest states;
this makes the evidence for the so-called adjoint string
breaking\footnote{See \cite{Schilling-99} 
  for a review on string breaking, and \cite{KF-03} for a
  recent related study in the $(2{+}1){-}d$ $SU(2)$ theory.} 
very difficult when using the method applied here, requiring a
prohibitive amount of statistics.
Indeed, both the $L=16$ and $L=12$ data seem to correspond
to the propagation of a string state with 
$\sigma_{\rm adj}\approx 0.145$, and therefore
$\sigma_{\rm adj}/\sigma\approx 2.1$, which is again 
rather close to the corresponding Casimir ratio $9/4$.

\bigskip
\noindent
{\bf Acknowledgments}

\noindent
We thank Misha Shifman for interesting discussions.



\begin{thebibliography}{99}

\bibitem{DFGO-96}
L. Del Debbio, M. Faber, J. Greensite, and \v{S}. Olejn\'ik,
Phys. Rev. D 53 (1996) 5891.

\bibitem{Deldar-00}
S. Deldar, Phys. Rev. D 62 (2000) 034509.

\bibitem{Bali-00}
G. Bali, Phys. Rev. D 62 (2000) 114503.

\bibitem{SS-00}
V. I. Schevchenko and Yu. A. Simonov,
Phys. Rev. Lett. 85 (2000) 1811.

\bibitem{LT-01}
B. Lucini and M. Teper,
Phys. Rev. D 64 (2001) 105019;
Phys. Lett. B 501 (2001) 128.

\bibitem{DPRV-02}
L. Del Debbio, H. Panagopoulos, P. Rossi, and E. Vicari, 
Phys. Rev. D 65 (2002) 021501;
JHEP 0201 (2002) 009.

\bibitem{Strassler-98}
M. J. Strassler, Prog. Theor. Phys. Suppl. {\bf 131}, 439
(1998) [e-print hep-th/9803009];
Nucl. Phys. (Proc. Suppl.) {\bf 73}, 120 (1999)
[e-print hep-lat/9810059].

\bibitem{AS-03a}
A. Armoni and M. Shifman, 
hep-th/0304127 Nucl. Phys. B, in press.

\bibitem{AS-03b}
A. Armoni and M. Shifman, hep-th/0307020.

\bibitem{DSST-85}
Ph. de Forcrand, G. Schierholz, H. Schneider, and M. Teper,
Phys. Lett. B 160 (1985) 137.

\bibitem{LSW-80}
M. L\"uscher, K. Symanzik, and P. Weisz,
Nucl. Phys. B 173 (1980) 365.

\bibitem{smearing}
M. Teper, 
Phys. Lett. B 183 (1987) 345.

M. Albanese et al. (APE Collaboration), 
Phys. Lett. B 192 (1987) 163.

\bibitem{Schilling-99}
K. Schilling, Nucl. Phys. Proc. Suppl. 83 (2000) 140
[hep-lat/9909152].

\bibitem{KF-03}
S. Kratochvila and Ph. de Forcrand,
hep-lat/0306011.

\end{thebibliography}
\end{document}